# On momentum conservation and thermionic emission cooling

Raseong Kim, Changwook Jeong, and Mark S. Lundstrom

Network for Computational Nanotechnology

Birck Nanotechnology Center

Purdue University, West Lafayette, IN 47907

**Abstract**

The question of whether relaxing momentum conservation can increase the performance of thermionic cooling device is examined. Both homojunctions and heterojunctions are considered. It is shown that for many cases, a non-conserved lateral momentum model overestimates the current. For the case of heterojunctions with a much heavier effective mass in the barrier and with a low barrier height, however, non-conservation of lateral momentum may increase the current. These results may be simply understood from the general principle that the current is limited by the location, well or barrier, with the smallest number of conducting channels. These results also show that within thermionic emission framework, the possibilities of increasing thermionic cooling by relaxing momentum conservation are limited. More generally, however, when the connection to the source is weak or in the presence of scattering, the situation may be different. Issues that deserve further study are identified.

## I. Introduction

Thermionic (TI) cooling is a method of refrigeration with the potential for high cooling power and efficiency.[1-3] As depicted in Fig. 1, it is based on thermionic emission over a potential barrier. When carriers with high energy (hot carriers) are injected over the barrier, the carrier distribution in the emitter region is out of equilibrium. To restore equilibrium, cold carriers move up and populate higher energy states by absorbing heat from the lattice. The result is that cooling occurs in the region before the emitter-barrier junction.[4,5] The purpose of this paper is to address the question of whether relaxing momentum conservation at the junction can increase the performance of TI cooling devices, as has been proposed.[6,7]

The main differences between TI cooling and the more conventional, thermoelectric (TE) cooling are the carrier transport mechanism and the operating regime.[8] In TI cooling, carrier transport from the well to the barrier is treated as ballistic, as is transport across the barrier (region II in Fig. 1) because the barrier thickness, $d$, is assumed to be shorter than the carrier mean-free-path, $\lambda$. The result is that no joule heating occurs in the channel.[4] In TE cooling, however, transport is assumed to be diffusive, and joule heating is considered in the heat balance.[9] In addition, while TE devices operate in the linear regime with small voltage and temperature differences, TI devices operate in the non-linear regime with high drain bias to eliminate the carrier injection from the drain and maximize the heat current injected from the source.[10]

Previous theoretical studies have compared the cooling performances of TI and TE devices.[8,11,12] It has been shown that for the same material, TE cooling is better because it gives higher maximum temperature difference, $\Delta T_{\max}$, than that obtained from TI cooling.[8] This result has been explained in terms of the "material parameter", $B$, where $B_{TE} > B_{TI}$.[11] These studies used two different sets of equations to model TI and TE devices. For TI devices, Richardson's equation[11] or its generalized version with Fermi-Dirac statistics[8] were used, and the Boltzmann transport equation (BTE)[13] was used to model TE devices. Using the Landauer approach,[14,15] however, it is possible to describe both TI and TE devices. In the diffusive limit the Landauer formalism describes TE devices and in the ballistic limit, TI devices.[16] As shown in Appendix A, we can calculate and compare $\Delta T_{max}$'s of TE and TI devices using the Landauer formula, and the result shows $B_{TE}/B_{TI} = \lambda/d > 1$, which is consistent with the result by Humphrey $et\ al.$[12] The physical explanation is that the short $d$ of TI device gives a large heat back-flow that limits $\Delta T_{max}$. The top-of-the-barrier model that we describe in the next section is closely related to the Landauer approach.

Although previous theoretical studies show that the cooling performance of TI devices is no better than that of TE devices, it has been suggested that non-conservation of lateral momentum may increase the number of electrons participating in the thermionic emission process and significantly improve the TI cooling performance.[6,7,10] The results to be presented in this paper, however, suggest that significant performance benefits are unlikely to be achieved. More generally, the results shed light on thermionic emission over barriers, a problem that is relevant to TI cooling but also important in other electronic devices such as Schottky barriers and metal-oxide-semiconductor field-effect transistors (MOSFETs).[17]

Our goals in this paper are to study the physics of thermionic emission across homo- and heterojunctions, to examine the concept of non-conserved lateral momentum, and to explore the interesting possibility of increasing TI device cooling performance by relaxing momentum conservation at the junction. The paper is organized as follows. In Sec. II, we compare two approaches to describe carrier injection over the barrier, a top-of-the-barrier model and a thermionic emission model. We also discuss the need for a general model for heterojunctions and introduce the concept of non-conserved lateral momentum. In Sec. III, the general theory of thermionic emission across heterojunctions is reviewed, and results are shown for homo- and heterojunctions. In Sec. IV, a simple physical interpretation is provided to explain the results in Sec. III. In Sec. V, we discuss the underlying physics of injection over a barrier, discuss the validity of the non-conservation lateral momentum model, and identify issues that deserve further attention. Conclusions follow in Sec. VI.

## II. Top-of-the-barrier and thermionic emission models

In this section, we compare two different approaches that describe carrier injection over a barrier, the top-of-the-barrier model[18] and the traditional thermionic emission model.[17] Both models assume ballistic transport. In the top-of-the-barrier model, the $E$–$k$ relation is considered on the barrier as shown in Fig. 2(a), and $+k_z$ states ($z$ is the transport direction) are filled according to the source Fermi level, $E_F$. In the thermionic emission model,[17] we focus on the source as shown in Fig. 2(b), and carriers with $k_z > k_b$ are injected from the well over the barrier. The value of $k_b$ is determined by the barrier height, $\phi_B$, as $k_b = \sqrt{2m^*\phi_B}/\hbar$, where $m^*$ is the carrier effective mass, and $\hbar$ is the reduced Planck's constant. Note that the conduction band edge, $E_C$, is assumed to be zero in the source region in Figs. 2(a)-(b). The condition $k_z > k_b$ implies that the lateral momentum is conserved during the emission process.

As discussed in Appendix B, it can be shown that the two approaches are equivalent for homojunctions, where $m^*$ is uniform in the source and the barrier region. As an example, Fig.

2(c) shows the $k$-space distribution of three-dimensional (3D) carriers that contribute to the current in the top-of-the-barrier picture (A) and thermionic emission picture (B) on the $k_z - k_x$ plane with $k_y = 0$. (Note that Fig. 2(c) also can be viewed as the $k$-space distribution of two-dimensional (2D) carriers.) Then for a single parabolic band, the ballistic electrical and heat currents are given as

$$I_{3D}/A = \frac{qm^*}{2\pi^2\hbar^3}\left(k_B T\right)^2 \mathscr{F}_1(\eta_F),\tag{1a}$$

$$I_{q,3D}/A = \frac{m^*}{2\pi^2\hbar^3}\left(k_B T\right)^3 \left(2\mathscr{F}_2(\eta_F) - \eta_F\mathscr{F}_1(\eta_F)\right),\tag{1b}$$

where $A$ is the cross-sectional area of the device, $q$ is the unit charge, $k_B$ is the Boltzmann constant, $T$ is the temperature, $\mathscr{F}_j$ is the Fermi-Dirac integral of order $j$,[19,20] and $\eta_F = \left(E_F - \phi_B\right)/k_B T$, which is the reduced Fermi level in the barrier region.

For heterojunction barriers, however, questions arise. For example, it is not clear which effective mass to use in eq. (1), the source mass, $m_1^*$, or the barrier mass, $m_2^*$. Questions also arise if we relax the assumption of conservation of lateral momentum inherent in the conventional thermionic emission approach. It has been suggested that non-conservation of lateral momentum may give higher emission current because carriers with $k > k_b$ are injected over the barrier while only those with $k_z > k_b$ are injected when the lateral momentum is conserved.[6,7] According to the top-of-the-barrier model, however, $+k_z$ states in the barrier are already in equilibrium with the source and no additional current is possible. In the next section, we review the general theory of thermionic emission across homo- and heterojunctions to address these questions and examine the validity of the top-of-the-barrier model and the non-conserved lateral momentum model.

## III. Thermionic emission across heterojunctions

We begin with a review of the general theory of thermionic emission across heterojunctions as presented by Wu and Yang.[21] It is assumed that $m^*$ changes abruptly at the junction interface.[21-23] Wu and Yang assume that the total energy, $E$, and the lateral momentum, $\hbar k_\perp$, are conserved

$$E_{\perp,1} + E_{\parallel,1} = E_{\perp,2} + E_{\parallel,2} + \phi_B \,, \tag{2a}$$

$$m_1^* E_{\perp,1} = m_2^* E_{\perp,2} \,, \tag{2b}$$

where $E_\perp$ and $E_\parallel$ are the kinetic energies along the lateral (perpendicular to transport) and longitudinal directions, and subscripts 1 and 2 denote source and the barrier regions, respectively. It can be shown that eq. (2) guarantees flux continuity across the barrier.[21,23] In this work, we use a semi-classical transmission for simplicity, so the transmission is 1 for carriers satisfying eq. (2) and 0 otherwise. Using a quantum mechanically computed transmission[21] would not change our conclusions. From now on, we call this approach the "conserved lateral momentum model."

In summary, we have three approaches to describe thermionic emission over the barrier: 1) the conserved lateral momentum (CLM) model, 2) the top-of-the-barrier (TOB) model, and 3) the non-conserved lateral momentum (NCLM) model. In the CLM model, total energy and lateral momentum are conserved as shown in eq. (2), and the theory applies generally for homo- and heterojunctions. In the TOB model, $+k_z$ states on the barrier are filled according to $E_F$ without considering the injection mechanism from the source. In the NCLM model, carriers with $k > k_b$ are injected from the source without considering the occupation of states on the barrier. Using these three approaches, we examine three cases: 1) homojunction with barrier, 2) heterojunction with no barrier, and 3) heterojunction with barrier. For heterojunctions, we consider two cases: i) $m_1^* > m_2^*$ and ii) $m_1^* < m_2^*$. The mathematics of these three cases is discussed in Appendix B; only the results are discussed below.

Results for a homojunction with a barrier are shown in Fig. 3. As discussed in Sec. II and depicted in Fig. 3(a), the CLM and TOB models are equivalent for homojunctions. For the NCLM model in Fig. 3(b), however, it is not clear how to map the $k$-states in the source to the barrier. Since all of the states in the barrier are already filled according to $E_F$ as shown in Fig. 3(a), it does not seem possible for a current in excess of that given by eq. (1) to flow.

In Fig. 4, we examine the case where $m^*$ changes abruptly but there is no potential barrier. For such cases, it is well known that the smaller $m^*$ determines the current,[24-26] and the electrical and heat currents are given from eq. (1) with lighter $m^*$. As shown in Fig. 4(a) when $m_1^* > m_2^*$, the current is determined by the states on the barrier, so the TOB model is valid and consistent with the CLM model, and the NCLM model (light red) overestimates the current. When $m_1^* < m_2^*$, however, the current is determined by the smaller $m_1^*$ in the source, so the TOB model (light blue) overestimates the current as shown in Fig. 4(b). Note that there is no distinction between the CLM and NCLM models in Fig. 4(b).

Next, we consider heterojunctions with potential barrier. When $m_1^* > m_2^*$, the CLM model is equivalent to the TOB model as shown in Fig. 5(a), and the total current is still determined by the lighter $m_2^*$ of the barrier. The current expression for 3D carriers is the same as eq. (1) with $m^* = m_2^*$. Note that the $k$-space distribution of carriers in the source that are able to surmount the barrier is different from the homojunction case that was shown in Fig. 3(a). (For details, see Appendix B.) Because the CLM model is consistent with the TOB model, the results from the NCLM model in Fig. 5(b) still cannot be mapped from the source to the barrier and would overestimate the current.

When $m_1^* < m_2^*$ with a potential barrier, there are two competing factors, the barrier height and the magnitude of the lighter effective mass. As discussed in the homojunction case and illustrated in Fig. 3(a), increasing $\phi_B$ tends to make the barrier states more dominant, while the lighter $m_1^*$ tends to make the source states more dominant as was illustrated in Fig. 4(a). We examine, therefore, two cases: i) $m_1^* <\sim m_2^*$ with high $\phi_B$ and ii) $m_1^* \ll m_2^*$ with low $\phi_B$. We expect that the states in the barrier will dominate in case i) while the source states will in case ii).

When $m_1^* <\sim m_2^*$ with high $\phi_B$, the CLM model is equivalent to the TOB model as shown in Fig. 6(a), and the 3D ballistic current is given as eq. (1) with $m^* = m_2^*$. It should be noted that the current is determined by the heavier mass, $m_2^*$, of the barrier, unlike the case with $\phi_B = 0$ in Fig. 4(a). Detailed expressions are shown in Appendix B. Because the current is determined by the states in the barrier, the NCLM model still cannot be mapped to the barrier and would overestimate the current as shown in Fig. 6(b).

The second case, $m_1^* \ll m_2^*$ with low $\phi_B$, is examined in Fig. 7. Note that the CLM model in Fig. 7(a) is different from both the NCLM model and the TOB model. (See Appendix B for details.) Note that as shown in Fig. 7(a), the states in the barrier are not completely filled by the source distribution function unlike other cases shown in Figs. 3(a), 5(a), and 6(a). There is, therefore, room for improving the emission current, and the maximum possible current is given by the NCLM model as shown in Fig. 7(b). The TOB model in Fig. 7(c) overestimates the current because it is larger than the maximum that can be supplied by the source, which is given by the NCLM model. In this case, it appears that the proposed increase in TI cooling by relaxing momentum conservation[6,7] could be achieved.

The possible improvement due to non-conservation of lateral momentum when $m_1^* \ll m_2^*$ with $\phi_B < \sim k_B T$ is calculated in Fig. 8. For a model 3D device with $m_1^* = 0.25 m_0$, $m_2^* = m_0$, and $\phi_B = 50$ meV, where $m_0$ is the free electron mass, the improvement of $I$ is about 18 % at $\eta_F = -1$ as shown in Fig. 8(a), and it is about 8 % for $I_q$ as shown in Fig. 8(b). The improvement is modest because the carrier distributions are already similar in the CLM and NCLM models as shown in Figs. 7(a)-(b). Note that the condition of $m_1^* \ll m_2^*$ with $\phi_B < \sim k_B T$ is not common for practical devices, and typical heterojunction devices[7] fall into the regime discussed in Figs. 5-6.

## IV. Conductance and minimum number of modes

The results in the previous section can be understood with a simple general rule. Given a number of conducting channels (or modes) in the source ($M_1(E)$) and the barrier ($M_2(E)$), the smaller one determines the total conductance.[27] As an example, we consider a 3D heterojunction where the numbers of modes increase linearly with $E$, and the slope is proportional to $m^*$ as[16]

$$M_1(E) = \frac{m_1^* E}{2\pi \hbar^2}, \tag{3a}$$

$$M_2(E) = \frac{m_2^* (E - \phi_B)}{2\pi \hbar^2}. \tag{3b}$$

Three different cases are considered in Fig. 9. In Fig. 9(a) where $m_1^* > m_2^*$, $M_1(E) < M_2(E)$ regardless of the value of $\phi_B$, so $M_2(E)$ determines the conductance. For heterojunctions with $m_1^* < m_2^*$, we consider two cases: 1) $m_1^* < \sim m_2^*$ with $\phi_B \gg k_B T$ as shown in Fig. 9(b) and 2) $m_1^* \ll m_2^*$ with $\phi_B < \sim k_B T$ as shown in Fig. 9(c). In Fig. 9(b), although $m_2^*$ in the barrier is heavier, due to the high $\phi_B$, $M_2(E)$ is smaller than $M_1(E)$ and determines the current. In Fig. 9(c) for case 2), however, $M_1(E) < M_2(E)$ because the much lighter $m_1^*$ in the source dominates despite the potential barrier, so it is the carrier injection from the source that limits the current. In this case, non-conservation of lateral momentum may help increase the emission current by maximizing the carrier injection from the source.

The results above are summarized in Table 1. The TOB model represents an upper limit to the possible current while the NCLM model represents the maximum current that could be supplied by the source if there were states in the barrier to accept them; the minimum of the two determines the current. In many cases, the TOB model gives correct results while the NCLM model overestimates the current. In cases where the number of modes in the source is smaller than that of the barrier ($m_1^* \ll m_2^*$ with low $\phi_B$), non-conservation of lateral momentum may increase the emission current.

## V. Discussion

In this section, we examine the underlying physics of the TOB and NCLM models and explore how the maximum emission current can be achieved. From now on, we mainly consider heterojunctions with a constant $m^*$. In the Landauer formula,[14,15] current on the barrier is expressed as

$$I = \frac{2q^2}{h} \int dE\, T(E) M(E) f(E) \qquad (4)$$

where $T(E)$ is the transmission, $M(E)$ is the number of modes, and $f(E)$ is the carrier distribution function. The maximum current is achieved for ballistic transport with $T(E) = 1$, and $M(E)$ is determined by device dimensionality and bandstructure.[16] The TOB model assumes that $f(E) = f_0(E)$, where $f_0(E)$ is the equilibrium Fermi-Dirac distribution of the source region.[18] As illustrated in Fig. 3 for the TOB model, non-conservation of lateral momentum cannot increase the current because the states on the barrier are already completely filled by the source distribution function in the CLM model.

For homojunctions or other cases in Table 1 where the TOB model is consistent with the CLM model, $f(E)$ must be larger than $f_0(E)$ to achieve the increase in current predicted by the NCLM model. It should be noted that for one-dimensional (1D) carriers, there is no distinction between the CLM and NCLM models because there are no transverse modes.[28] For 2D and 3D carriers, if $f(E)$ is assumed to be uniform along the angular directions, then we can show that $f_{2D}(E) = \sqrt{E/(E - \phi_B)} \times f_0(E)$ and $f_{3D}(E) = E/(E - \phi_B) \times f_0(E)$ for homojunctions. (See Appendix B for details.) Previous theoretical works[6,7] that report an order of magnitude increase of conductivity in the NCLM model implicitly assume that the states in the barrier are populated above their equilibrium value. This does not appear to be physically possible.

It has been suggested that in a non-planar heterostructure where the translational invariance is broken, the lateral momentum is not conserved, and the emission current may increase.[6,7,29,30] Monte Carlo simulations including inelastic scattering processes[29,30] have shown that adjusting the depth and the width of the zigzagged interface structure enhances the emission by a factor of up to 2. Note, however, that the lateral momentum is conserved at each local interface, so we may interpret this enhancement due to the increased effective area. The current is not directly proportional to the total interface area[31] because the carriers may re-enter the emitter in the zigzagged structure.[29,30] To better understand the physics of carrier emission and explore possibilities to increase the emission current, we need to study the effect of carrier ballisticity on the emission enhancement and optimize the non-planar structures[29,30,32] to maximize the enhancement while not decreasing the carrier mobility.[6]

It has been shown that non-conserved lateral momentum is essential to interpret the experimental results of Ballistic Electron Emission Microscopy (BEEM) for non-epitaxial metal-semiconductor interfaces.[33-35] In BEEM measurements, carriers with small lateral momentum are predominantly injected.[33] The conserved lateral momentum model does not provide physical interpretation for the experiment where valleys with zero lateral momentum are not preferentially populated as would be expected if lateral momentum were conserved.[33] The observed significant current for the valleys with non-zero lateral momentum indicates that additional lateral momentum is provided by scattering at the non-epitaxial interface.[35] The BEEM measurement results and the theories of non-conservation of lateral momentum used to explain them have motivated the idea that non-conservation of lateral momentum might similarly enhance the emission current and cooling performance of TI devices.[6] The problems are, however, quite different. The critical difference between BEEM experiments and TI devices is that the carrier reservoirs are different. In TI devices, the source should be designed to act as closely as possible to an ideal Landauer reservoir,[36] where the equilibrium distribution is maintained by a high carrier density, high number of modes, and high scattering rates. Such a contact can provide carriers with all possible $k$'s with any given $E$. For such cases, the NCLM model may be unphysical or give only moderate improvements depending on $m^*$ and $\phi_B$ as discussed in previous sections. In BEEM experiments, however, the reservoir is far from ideal because the lateral momentum of injected carriers is predominantly zero and the connection to the source is weak. For such conditions, non-conservation of lateral momentum may help increase the emission current by shuffling the momentum distribution of carriers and performing the role of scattering in the ideal, Landauer contact. Although non-conservation of lateral momentum may help in such cases, the maximum current can never exceed the ballistic limit, which is determined by the minimum number of modes as summarized in Table 1.

Finally, we should mention that there are a number of other issues that deserve consideration. We have assumed a ballistic (thermionic emission) model in which all of the scattering occurs in the Landauer contacts. In practice, scattering will occur throughout the structure. In the well region before the barrier, scattering may reduce the current below the thermionic emission value. A similar problem, transport in Schottky barriers, was considered by Bethe[37] and by Berz.[38] Fischetti *et al.* have discussed the role of source starvation[39] in nanoscale MOSFETs. Here the idea is that the longitudinal momentum states that are injected over the barrier can become depleted if scattering in the well impeded the connection to the Landauer contact, which could replenish these states. Paradoxically, momentum randomizing scattering in the well could help by replenishing these longitudinal $k$-states. Of course, it would also increase the Joule losses, and we do not believe that cooling powers above the ballistic limit discussed in previous sections could be achieved. Fischetti *et al*. also discussed "downstream" effects[40] – scattering in the barrier itself and in the well beyond the barrier. Although they are beyond the scope of this paper, more quantitative studies of the effect of scattering on TI devices will be essential to understand the physics and performance limits of such devices.

## VI. Conclusions

In this paper, we studied the physics of thermionic emission across homo- and heterojunctions to explore the possibilities to increase the emission current and the cooling performance of TI cooling devices. We showed that the TOB model[18] is consistent with the thermionic emission model with conserved lateral momentum[21] for homojunctions, heterojunctions with heavier effective mass in the source, and heterojunctions with heavier effective mass in the barrier region and high barrier height, $\phi_B$. For such cases, the NCLM model[6] is not consistent with the TOB model and, we believe, overestimates the current that is possible. For heterojunctions with much heavier $m^*$ in the barrier with low $\phi_B$, however, we note that non-conservation of lateral momentum may increase the current because there are unfilled states in the barrier region when the lateral momentum is conserved. These results can be explained by a simple general rule that given a number of modes in the source and the barrier, the overall conductance is determined by the minimum of the two.[27] These results show that within thermionic emission framework, the possibilities of increasing thermionic cooling by relaxing momentum conservation are limited. Finally, we note that in the presence of scattering, the role of momentum randomization is an open question. For real TI cooling devices, as opposed to the ballistic devices connected to ideal, Landauer reservoirs considered here, momentum randomizing scattering in the well may enhance performance.

**Acknowledgments**

This work was supported by the Semiconductor Research Corporation. Computational support was provided by the Network for Computational Nanotechnology, supported by the National Science Foundation under Cooperative Agreement No. EEC-0634750. It is a pleasure to acknowledge helpful discussions with L. Siddiqui and Prof. S. Datta at Purdue University, Prof. A. Shakouri at the University of California at Santa Cruz, Dr. S. Jin at Synopsys Inc., and Prof. M. V. Fischetti at the University of Massachusetts at Amherst.

**Appendix A: Mathematical formulations of TE and TI devices**

According to the Landauer formalism,[14,15] the electrical and heat currents are expressed in terms of $T(E)$ and $M(E)$, and the formalism of Kim *et al*. applies to both TI and TE devices.[16] For TI devices (ballistic transport), $T(E) = 1$, and for TE devices (diffusive transport), $T(E) \approx \lambda(E)/L$, where $\lambda(E)$ is the $E$-dependent mean-free-path for backscattering, and $L$ is the length of the conductor. For several common scattering mechanisms, $\lambda(E)$ can be expressed in power law form as $\lambda(E) = \lambda_0 \left( E(p)/k_B T \right)^s$, where $\lambda_0$ is a constant, $E(p)$ is the kinetic energy, and $s$ is the characteristic exponent.[41]

In the linear regime where TE devices operate, transport coefficients such as conductance, $G$, and Seebeck coefficient, $S$, can be obtained in terms of $T(E)$ and $M(E)$.[16] The efficiency of TE devices is related to the figure of merit, $ZT = S^2 GT/K$,[9] where $K$ is the thermal conductance, which is the sum of the electronic contribution and the lattice thermal conductance, $K_l$. The relation between $ZT$ and $\Delta T_{\max}$ is $ZT \equiv 2\Delta T_{\max}/T$.[8] Then $ZT$ can be calculated using the expressions for $S$, $G$, and $K$ from the Landauer formula.[16] For a 3D single parabolic band, for example, we find

$$ZT = \frac{\left( (s+2)\mathscr{F}_{s+1}(\eta_F)/\mathscr{F}_s(\eta_F) - \eta_F \right)^2}{(s+3)(s+2)\dfrac{\mathscr{F}_{s+2}(\eta_F)}{\mathscr{F}_s(\eta_F)} - (s+2)^2\dfrac{\mathscr{F}_{s+1}^2(\eta_F)}{\mathscr{F}_s^2(\eta_F)} + \dfrac{2\pi^2\hbar^3\kappa_l}{m^* k_B (k_B T)^2 \lambda_0 \Gamma(s+2)\mathscr{F}_s(\eta_F)}}, \qquad (A1)$$

where $\Gamma$ is the Gamma function. We can also define the material parameter for TE device, $B_{TE}$,[11,12] as $B_{TE} \equiv m^* k_B (k_B T)^2 \lambda_0/2\pi^2\hbar^3\kappa_l$, where $K_l = A\kappa_l/d$ with cross-sectional area $A$.

TI devices operate in the ballistic, non-linear regime. To calculate $\Delta T_{max}$ for a 3D single parabolic band, for example, we solve the heat balance equation, $I_q - A\kappa_l \Delta T_{max}/d = 0$.[8] Then we obtain

$$\frac{\Delta T_{max}}{T} = \frac{d}{\kappa_l}\frac{4\pi q m^* k_B^2}{h^3}\frac{k_B T^2}{q}\left(2\mathscr{F}_2(\eta_F) - \eta_F\mathscr{F}_1(\eta_F)\right) \equiv B_{TI}\left(2\mathscr{F}_2(\eta_F) - \eta_F\mathscr{F}_1(\eta_F)\right), \qquad (A2)$$

where $B_{TI}$ is the material parameter for TI devices.[11,12] By comparing eqs. (A1) and (A2), we obtain $B_{TE}/B_{TI} = \lambda_0/d$, which is consistent with the previously reported result.[12]

## Appendix B: Mathematics of the TOB and Wu-Yang Models

In the TOB model, the ballistic $I$ and $I_q$ for 3D carriers are calculated as

$$I_{3D,TOB}/A = \frac{q}{4\pi^3}\int_0^\infty dk\,k^2\int_0^{\pi/2}d\theta\sin\theta\int_0^{2\pi}d\phi\frac{\hbar k}{m_2^*}\cos\theta f_0 = \frac{q m_2^*}{2\pi^2\hbar^3}\left(k_B T\right)^2\mathscr{F}_1(\eta_F), \qquad (B1a)$$

$$\begin{aligned}I_{q,3D,TOB}/A &= \frac{1}{4\pi^3}\int_0^\infty dk\,k^2\int_0^{\pi/2}d\theta\sin\theta\int_0^{2\pi}d\phi\left(E - E_F\right)\frac{\hbar k}{m_2^*}\cos\theta f_0 \\ &= \frac{m_2^*}{2\pi^2\hbar^3}\left(k_B T\right)^3\left(2\mathscr{F}_2(\eta_F) - \eta_F\mathscr{F}_1(\eta_F)\right)\end{aligned}. \qquad (B1b)$$

Expressions for 1D and 2D carriers can be obtained in a similar way. We split the CLM model[21] into two cases, $m_1^* > m_2^*$ and $m_1^* < m_2^*$. For $m_1^* > m_2^*$, the case in Fig. 5(a), the results are

$$\begin{aligned}I_{3D,CLM}/A &= \frac{q}{4\pi^3}\int_{\sqrt{2m_1^*\phi_B}/\hbar}^\infty dk\,k^2\int_0^{\sin^{-1}\sqrt{\frac{m_2^*}{m_1^*}\left(1 - \frac{2m_1^*\phi_B}{\hbar^2 k^2}\right)}}d\theta\sin\theta\int_0^{2\pi}d\phi\frac{\hbar k}{m_1^*}\cos\theta f_0, \\ &= \frac{q m_2^*}{2\pi^2\hbar^3}\left(k_B T\right)^2\mathscr{F}_1(\eta_F)\end{aligned} \qquad (B2a)$$

$$I_{q,3D,CLM}/A = \frac{1}{4\pi^3} \int\limits_{\sqrt{2m_1^*\phi_B}/\hbar}^{\infty} dk k^2 \int\limits_0^{\sin^{-1}\sqrt{\frac{m_2^*}{m_1^*}\left(1-\frac{2m_1^*\phi_B}{\hbar^2 k^2}\right)}} d\theta \sin\theta \int\limits_0^{2\pi} d\phi \left(E-E_F\right) \frac{\hbar k}{m_1^*}\cos\theta f_0 , \tag{B2b}$$

$$= \frac{m_2^*}{2\pi^2\hbar^3}\left(k_B T\right)^3 \left(2\mathscr{F}_2(\eta_F) - \eta_F\mathscr{F}_1(\eta_F)\right)$$

and we note that the results of eq. (B2) are the same as those from eq. (B1). In Fig. 5(a), the hyperbola (A) that maps onto the $k_x - k_y$ plane with $k_z = 0$ on the barrier is expressed as

$$\frac{\hbar^2 k_z^2}{2m_1^*} - \left(\frac{m_1^*}{m_2^*}-1\right)\frac{\hbar^2\left(k_x^2+k_y^2\right)}{2m_1^*} = \phi_B . \tag{B3}$$

For $m_1^* < m_2^*$, the case in Figs. 6(a) and 7(a), $I$ and $I_q$ become

$$I_{3D,CLM}/A = \frac{q}{4\pi^3} \int\limits_{\frac{\sqrt{2m_1^*\phi_B}}{\hbar}}^{\frac{\sqrt{2m_1^*\phi_B}}{\hbar}\sqrt{\frac{m_2^*}{m_2^*-m_1^*}}} dk k^2 \int\limits_0^{\sin^{-1}\sqrt{\frac{m_2^*}{m_1^*}\left(1-\frac{2m_1^*\phi_B}{\hbar^2 k^2}\right)}} d\theta \sin\theta \int\limits_0^{2\pi} d\phi \frac{\hbar k}{m_1^*}\cos\theta f$$

$$+ \frac{q}{4\pi^3} \int\limits_{\frac{\sqrt{2m_1^*\phi_B}}{\hbar}\sqrt{\frac{m_2^*}{m_2^*-m_1^*}}}^{\infty} dk k^2 \int\limits_0^{\pi/2} d\theta \sin\theta \int\limits_0^{2\pi} d\phi \frac{\hbar k}{m_1^*}\cos\theta f \tag{B4a}$$

$$= \frac{q m_1^*}{2\pi^2\hbar^3}\left(k_B T\right)^2 \int\limits_0^{\frac{m_1^*}{m_2^*-m_1^*}\frac{\phi_B}{k_B T}} dx \frac{x}{1+e^{x-\eta_F}} + \frac{q m_1^*}{2\pi^2\hbar^3}\left(k_B T\right)^2 \int\limits_{\frac{m_1^*}{m_2^*-m_1^*}\frac{\phi_B}{k_B T}}^{\infty} dx \frac{x+\phi_B/k_B T}{1+e^{x-\eta_F}}$$

$$I_{q,3D,CLM}/A = \frac{1}{4\pi^3} \int\limits_{\sqrt{2m_1^*\phi_B}/\hbar}^{\frac{\sqrt{2m_1^*\phi_B}}{\hbar}\sqrt{\frac{m_2^*}{m_2^*-m_1^*}}} dk k^2 \int\limits_0^{\sin^{-1}\sqrt{\frac{m_2^*}{m_1^*}\left(1-\frac{2m_1^*\phi_B}{\hbar^2 k^2}\right)}} d\theta \sin\theta \int\limits_0^{2\pi} d\phi \left(E-E_F\right) \frac{\hbar k}{m_1^*}\cos\theta f$$

$$+ \frac{1}{4\pi^3} \int\limits_{\frac{\sqrt{2m_1^*\phi_B}}{\hbar}\sqrt{\frac{m_2^*}{m_2^*-m_1^*}}}^{\infty} dk k^2 \int\limits_0^{\pi/2} d\theta \sin\theta \int\limits_0^{2\pi} d\phi \left(E-E_F\right) \frac{\hbar k}{m_1^*}\cos\theta f \tag{B4b}$$

$$= \frac{m_2^*}{2\pi^2\hbar^3}\left(k_B T\right)^3 \int\limits_0^{\frac{m_1^*}{m_2^*-m_1^*}\phi_B} dx \frac{\left(x-\eta_F\right)x}{1+e^{x-\eta_F}} + \frac{m_1^*}{2\pi^2\hbar^3}\left(k_B T\right)^3 \int\limits_{\frac{m_1^*}{m_2^*-m_1^*}\phi_B}^{\infty} dx \frac{\left(x-\eta_F\right)\left(x+\phi_B/k_B T\right)}{1+e^{x-\eta_F}}$$

As $m_1^*/\left(m_2^*-m_1^*\right)\times\phi_B/k_BT\to\infty$ ( $m_1^*<\sim m_2^*$ or $\phi_B\gg k_BT$ ), we note that eq. (B4) approaches to eq. (B1), and the model becomes equivalent to the TOB model as shown in Fig. 6(a). As $m_1^*/\left(m_2^*-m_1^*\right)\times\phi_B/k_BT\to 0$ ( $m_1^*\ll m_2^*$ with low $\phi_B$ ), however, eq. (B4) is different from the TOB model as shown in Fig. 7(a). In Figs. 6(a) and 7(a), the ellipsoid (A) that maps onto the $k_x-k_y$ plane with $k_z=0$ line on the barrier is expressed as eq. (B3). In Fig. 7(a), the $k_x-k_y$ plane with $k_z=0$ in the source maps onto a hyperbola (B') on the barrier, which is given as

$$-\frac{\hbar^2 k_z^2}{2m_2^*}+\left(1-\frac{m_1^*}{m_2^*}\right)\frac{\hbar^2\left(k_x^2+k_y^2\right)}{2m_1^*}=\phi_B \ .\qquad(B5)$$

In the NCLM model, $I$ and $I_q$ for 3D carriers are

$$I_{3D,NCLM}\big/A=\frac{1}{4\pi^3}\int\limits_{\sqrt{2m_1^*\phi_B}/\hbar}^{\infty}dk\,k^2\int\limits_0^{\pi/2}d\theta\sin\theta\int\limits_0^{2\pi}d\phi\left(E-E_F\right)\frac{\hbar k}{m_1^*}\cos\theta f_0$$
$$=\frac{qm_1^*}{2\pi^2\hbar^3}\left(k_BT\right)^2\left(\mathscr{F}_1(\eta_F)+\frac{\phi_B}{k_BT}\mathscr{F}_0(\eta_F)\right)\qquad(B6a)$$

$$I_{q,3D,NCLM}\big/A=\frac{1}{4\pi^3}\int\limits_{\sqrt{2m_1^*\phi_B}/\hbar}^{\infty}dk\,k^2\int\limits_0^{\pi/2}d\theta\sin\theta\int\limits_0^{2\pi}d\phi\left(E-E_F\right)\frac{\hbar k}{m_1^*}\cos\theta f_0$$
$$=\frac{m_1^*}{2\pi^2\hbar^3}\left(k_BT\right)^3\left(2\mathscr{F}_2(\eta_F)+\left(\frac{\phi_B}{k_BT}-\eta_F\right)\mathscr{F}_1(\eta_F)-\frac{\phi_B}{k_BT}\eta_F\mathscr{F}_0(\eta_F)\right),\qquad(B6b)$$

and by comparing eqs. (B1) and (B6), we can show that $f\left(E\right)$ on the barrier should satisfy $f\left(E\right)=E/\left(E-\phi_B\right)\times f_0\left(E\right)$ to be consistent with the NCLM model for 3D homojunctions. We can also obtain the corresponding relation for 2D carriers.

**Table Captions**

Table 1. Summary of the general rule that determines the ballistic current across heterojunctions. The TOB model represents an upper limit to the possible current while the NCLM model represents the maximum current that could be supplied by the source, and the minimum of the two determines the total current.

**Figure Captions**

Fig. 1. Potential profile of TI device with drain bias $V_D$ . When hot carriers are injected over the barrier, cold carriers absorb heat from the lattice and populate higher energy states to restore the equilibrium distribution, $f_0(E)$, and cooling occurs in the region before the emitter-barrier junction. Carrier transport is assumed to be ballistic in region II while the equilibrium distribution is maintained in the diffusive regions, I and III.

Fig. 2. Two approaches that describe thermionic emission across homojunctions, (a) top-of-the-barrier model and (b) thermionic emission model, and (c) $k$-space distribution of 3D carriers contributing to the current shown on the $k_z - k_x$ plane with $k_y = 0$. (a) $E$–$k$ relation is considered in the barrier, and the $+k_z$ states are filled according to $E_F$ . (b) $E$–$k$ relation is considered in the source, and carriers with $k_z > k_b$ are injected from the well over the barrier.

Fig. 3. Results for a homojunction with $\phi_B$ . (a) The CLM model becomes equivalent to the TOB model. (b) It is not clear how the NCLM model can be described in the barrier because states on the barrier are already filled according to $E_F$ .

Fig. 4. Results for a heterojunction with $\phi_B = 0$. The smaller $m^*$ determines the current. (a) When $m_1^* > m_2^*$, the current is determined by the barrier states. (b) When $m_1^* < m_2^*$, the current is determined by the source states.

Fig. 5. Results for a heterojunction with $m_1^* > m_2^*$ and $\phi_B$. (a) The CLM model is equivalent to the TOB model. The hyperbola (A) in the source (eq. (B3)) is mapped onto the $k_x - k_y$ plane with $k_z = 0$ (A') on the barrier. (b) The results from the NCLM model overestimate the current and cannot be mapped to the barrier.

Fig. 6. Results for a heterojunction with $m_1^* <\sim m_2^*$ and high $\phi_B$. (a) The CLM model is equivalent to the TOB model. The ellipsoid (A) in the source (eq. (B3)) is mapped onto the $k_x - k_y$ plane with $k_z = 0$ (A') on the barrier. (b) The results from the NCLM model overestimate the current and cannot be mapped to the barrier.

Fig. 7. Results for a heterojunction with $m_1^* \ll m_2^*$ and low $\phi_B$. (a) The CLM model is different from both the NCLM model and the TOB model. The ellipsoid (A) in the source (eq. (B3)) is mapped onto the $k_x - k_y$ plane with $k_z = 0$ (A') on the barrier, and the $k_x - k_y$ plane with $k_z = 0$ (B) in the source is mapped onto the hyperbola (B') on the barrier (eq. (B5)). (b) The maximum possible current is given by the NCLM model. (c) The TOB model overestimates the current.

Fig. 8. Possible improvement over the CLM model (blue circle) due to NCLM (red cross) for a 3D model device with $m_1^* = 0.25m_0$, $m_2^* = m_0$, and $\phi_B = 50$ meV. (a) The improvement of $I$ is about 18 % at $\eta_F = -1$ (eq. (B6a)). The TOB model (green rectangle, eq.(1a)) overestimates the current because it is larger than that from the NCLM model. (b) The improvement of $I_q$ is about 8 % at $\eta_F = -1$ (eq. (B6b)).

Fig. 9. General rule to determine the emission current across heterojunctions. For $M_1(E)$ and $M_2(E)$, the smaller one determines the current. (a) When $m_1^* > m_2^*$, $M_1(E) < M_2(E)$ regardless of the value of $\phi_B$. (b) When $m_1^* <\sim m_2^*$ with $\phi_B \gg k_B T$, although $m_2^*$ is heavier, $M_1(E) > M_2(E)$ due to the high $\phi_B$. (c) When $m_1^* \ll m_2^*$ with $\phi_B <\sim k_B T$, $M_1(E) < M_2(E)$ because the much lighter $m_1^*$ dominates despite $\phi_B$.

**Table 1**

| | | Conserved lateral momentum | Top-of-the barrier | Non-conserved lateral momentum |
|---|---|---|---|---|
| homojunction | | correct | correct | incorrect |
| heterojunction $m_1^* > m_2^*$ | | correct | correct | incorrect |
| heterojunction $m_1^* < m_2^*$ | $m_1^* <\sim m_2^*$ $\phi_B \gg k_B T$ | correct | correct | incorrect |
| | $m_1^* \ll m_2^*$ $\phi_B <\sim k_B T$ | correct | incorrect | possible |

**Figure 1**

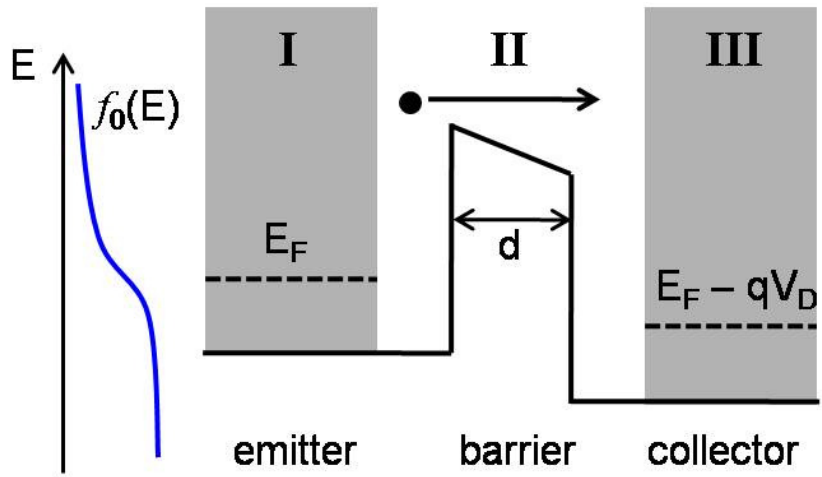

Figure 2

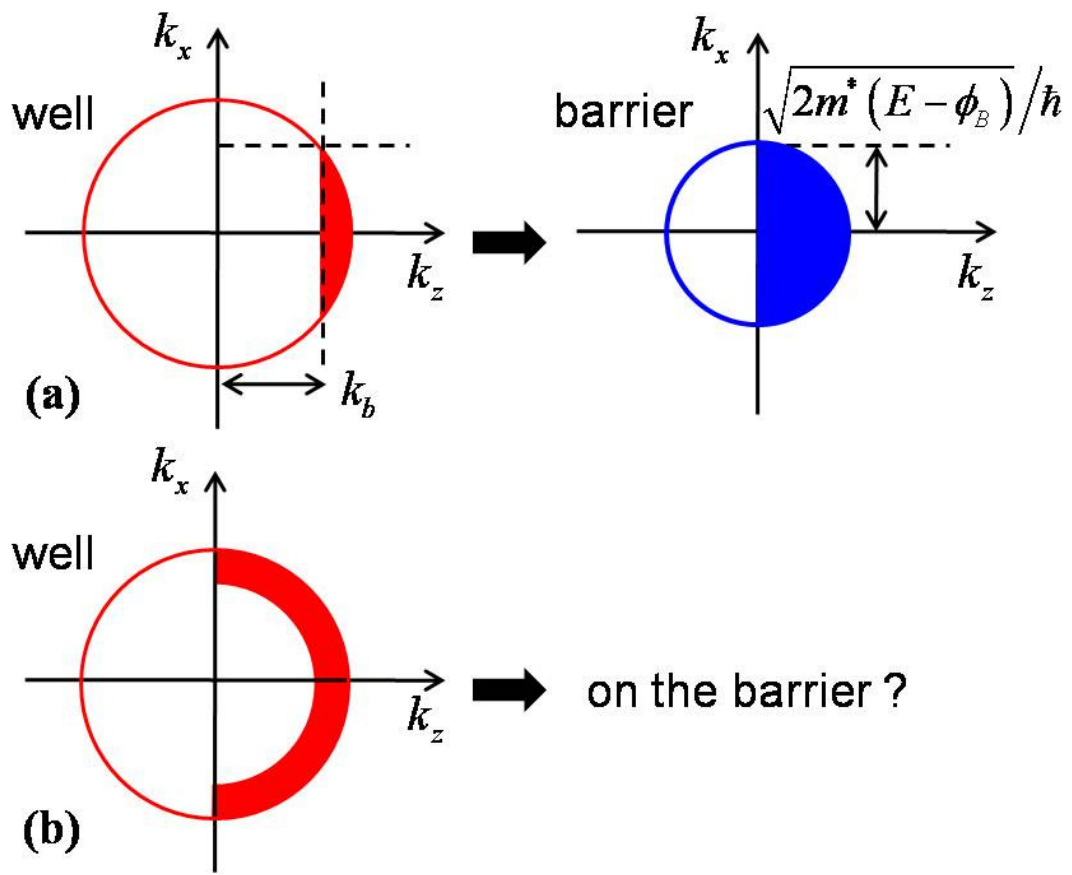

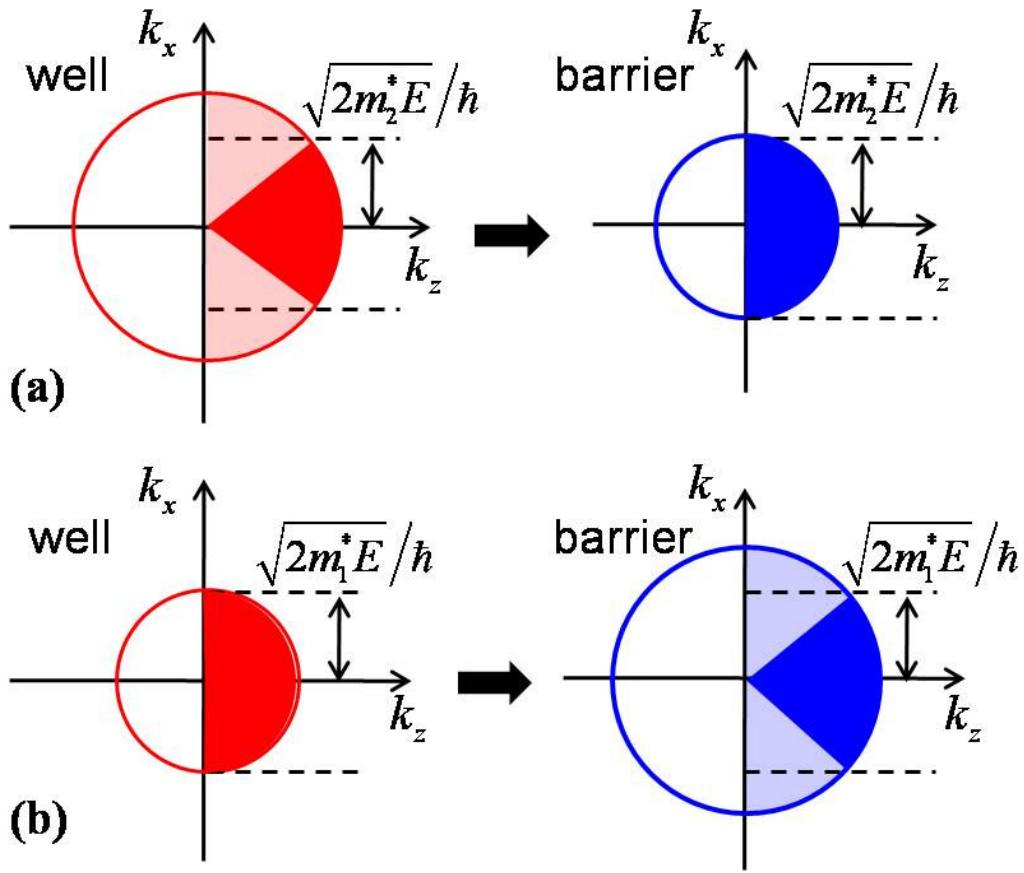

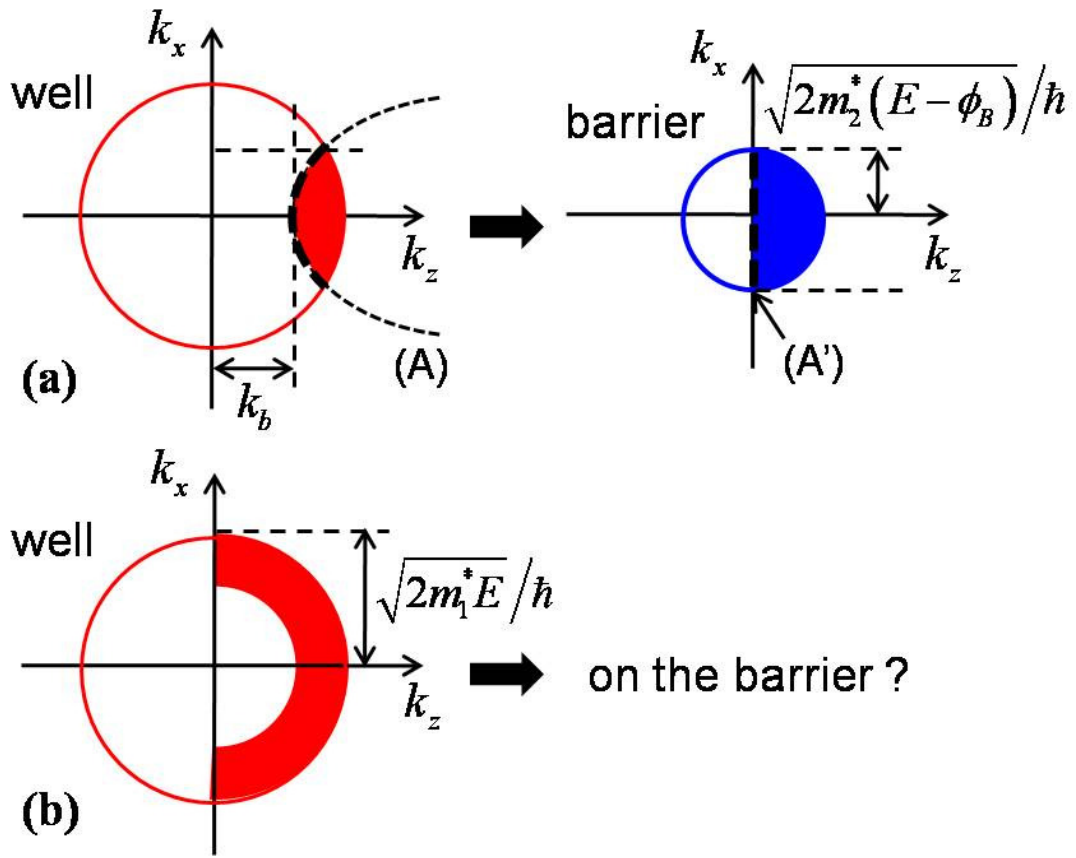

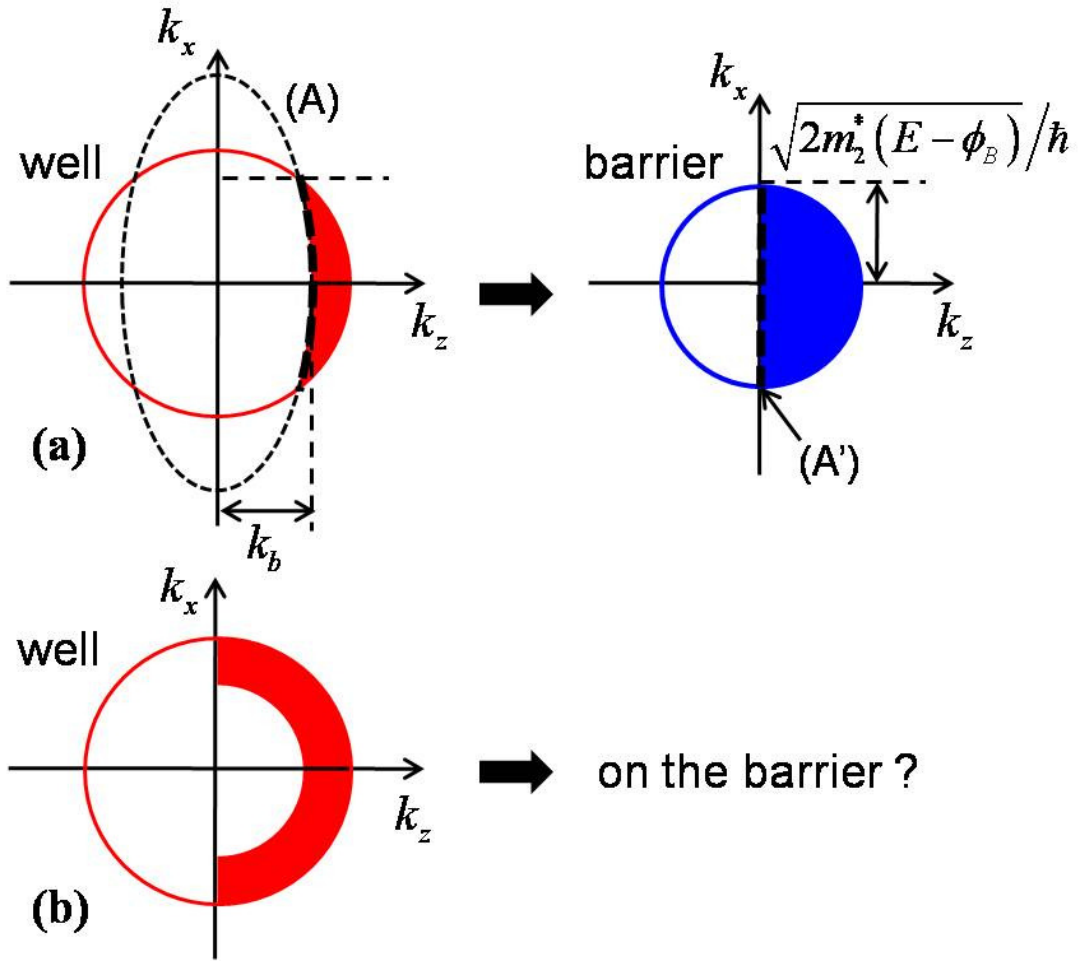

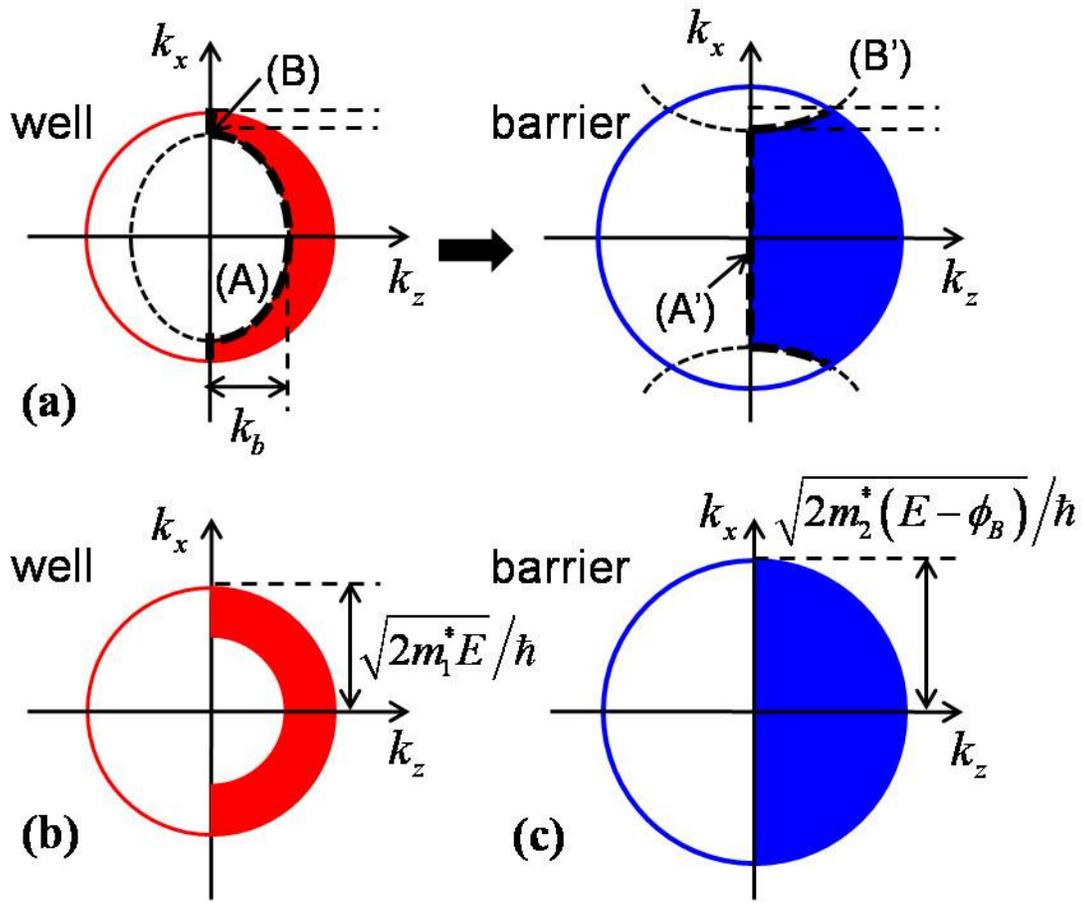

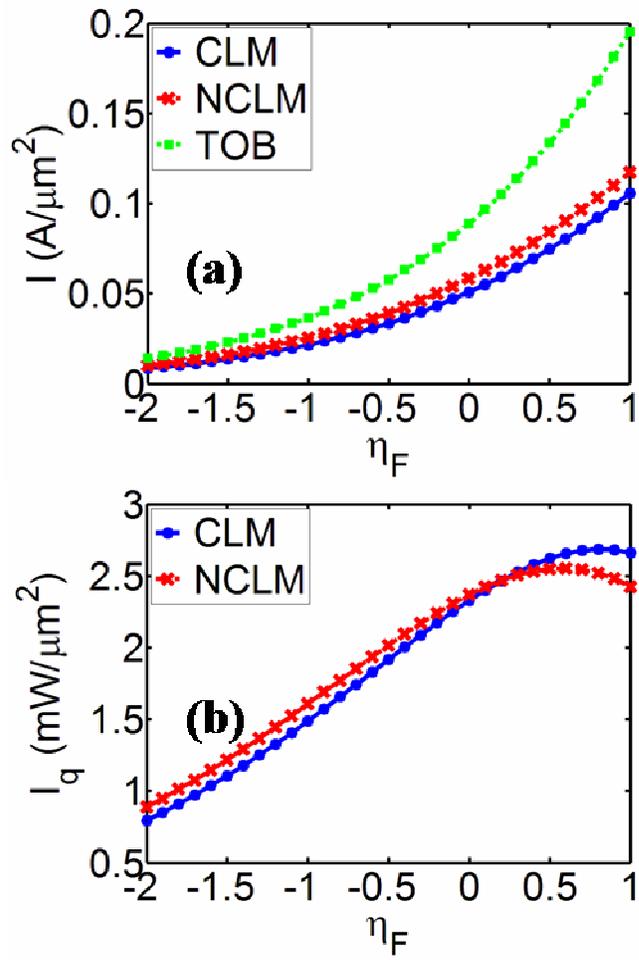

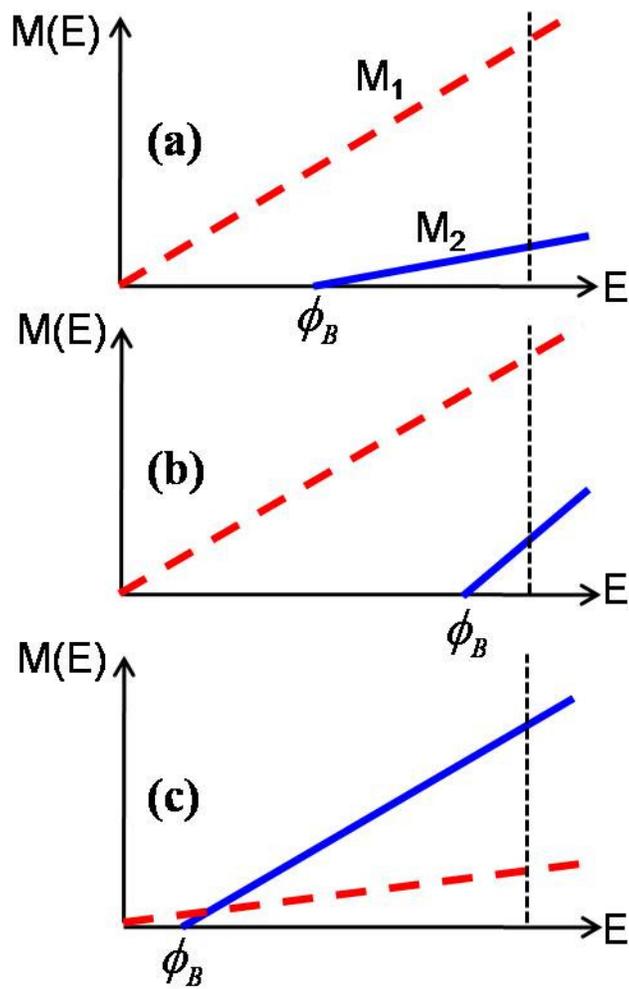